# Document classification methods

Madjid Khalilian, Shiva Hassanzadeh

Islamic Azad University, karaj branch

khalilian@kiau.ac.ir, mshh_84@yahoo.com

**Abstract**

Information on different fields which are collected by users requires appropriate management and organization to be structured in a standard way and retrieved fast and more easily. Document classification is a conventional method to separate text based on their subjects among scientific text, web pages and digital library. Different methods and techniques are proposed for document classifications that have advantages and deficiencies. In this paper, several unsupervised and supervised document classification methods are studied and compared.

## 1. Introduction

With increasing growth of communication, a great volume of data is subjected to the users. Thus, need to store, organize and manage this great volume of data is necessary more than ever. The task of a classification system is that designs a standard model suitable for existing data not only to provide structural consistency, but also to facilitate retrieving. Document classification is dividing document to some similar groups. In each group high degree of similarity is existed while similarity among document belongs to different category should be as small as possible. Classification can be divided to two categories: supervised classification or with model and unsupervised classification or without model. In supervised classification, labeled documents are group into pre-determined classes. It means that, a model can be constructed according to existing sample based on which unlabeled data assigned to their respective categories. In supervised, external measures are used to evaluate the results based on class label. External measures are based on pre-determined structures that reflect previous information of data. These measures used as a standard support for validity of classification results. Among these measures, F-measure is obtained as ratio of recall and precision variables. Precision variable is defined as the ratio of retrieved and related documents to the retrieved documents. In fact, this variable determines the degree of accuracy of results produced by the classification algorithm. Also, recall variable is characterized as ratio of related and retrieved documents to related documents.

In unsupervised classification, the documents are unlabeled and classified into clusters based on distance measure. Clustering is an indirect data mining process that means to group similar objects in the clusters and it seeks to explore data structure through similarity and dissimilarity between them. In this method there are no predetermined groups and the focus is on the data which are similar together to better identify the behaviors through finding these similarity and make decision based on better recognition. External and internal measures are defined to evaluate the quality of the clusters. Internal measure evaluates clustering structure directly from original data. These measures include error square which indicates the error rate of cluster data from cluster center and average silhouette value which is a combination of compactness and separateness for clusters. Also, external measures such as purity, precision, recall and F-measure are used. There are several methods of clustering [1] which classified as indicated in figure1 [2].

Partitioning clustering is one kind of clustering algorithms which firstly divided dataset to some parts and then evaluate them using some classification measures and applied some changes on initial classification if necessary. K-means algorithm is one of the most common partitioning clustering that based on iteration. This algorithm is on unsupervised iterative algorithm in which dataset is divided to k cluster and data points are assigned to these clusters randomly. Then for each point, the distance from it to cluster is calculated and it is assigned to the nearest cluster. These steps are repeated until there is no point to change its position. Some characteristics of this algorithm include:

- Always k cluster is existed.
- Always there is at least one point in each cluster.
- Clusters are not in hierarchical order and dose not overlaps each other.
- Each member of cluster has the least distance from cluster center.

This algorithm despite of some advantages such as simplicity, easy to implementation, high speed and begin suitable for large dataset, has some challenges. Shortcomings of this algorithm are determining the number of clusters k, sensitivity to noisy data and outliers, dependency of final result to initial centers of clusters and number of clusters, and getting stuck in local optimum.

In hierarchical method, clustering has a structure similar to a tree and it is usually in two agglomerative (down-to-up) and divisive (up-to-down). In agglomerative clustering, clusters with one data are considered at the beginning (at first, the number of cluster equal to the number of data) in each step two or more suitable clusters are merged together and constructed a new cluster. From among several kinds of conventional agglomerative hierarchical algorithms, we can point to single-link, complete-link and average link. In divisive clustering; the process begins with one cluster. This cluster is divided to two or more clusters recursively, and process continues in the same way. A terminating condition is required for both algorithms that often include reaching to k cluster.

**Comparison between hierarchical and partitioning algorithms**

- The hierarchical methods require an appropriate parameter to stop the algorithms.
- Time complexity of the algorithm is $O(N^2)$ (N is the number of sample) which is not efficient with increase of data mass.
- Assignment in one step is not convertible in next steps.
- Partitioning methods generally are dependent in initialization and the result of clustering will be different by changing the initial value.
- Partitioning methods are suitable with large mass of data.
- The numbers of clusters have to be determined by user.
- Convertible in initial structure of cluster.

So far, several methods and algorithms have been proposed and applied for classified text documents. All these algorithms can be divided to three general categories: algorithms classifying document based on their structure or their content or combination of document content and structure that have their special advantages and disadvantages. What common in document classification are the preprocessing documents during which text features are more important than the other have to be extracted to keep text document in a vector form and it can be possible to find similar vectors easily by a measure of cosine distance. In order to vector presentation of the data, vector space model is used and then classification process is applied on them. Remain of this paper is organized as follow: in section 2, some of unsupervised documents classification is studied. In section 3, some supervised classification methods are evaluated and in section 4 proposed algorithms are discussed and concluded.

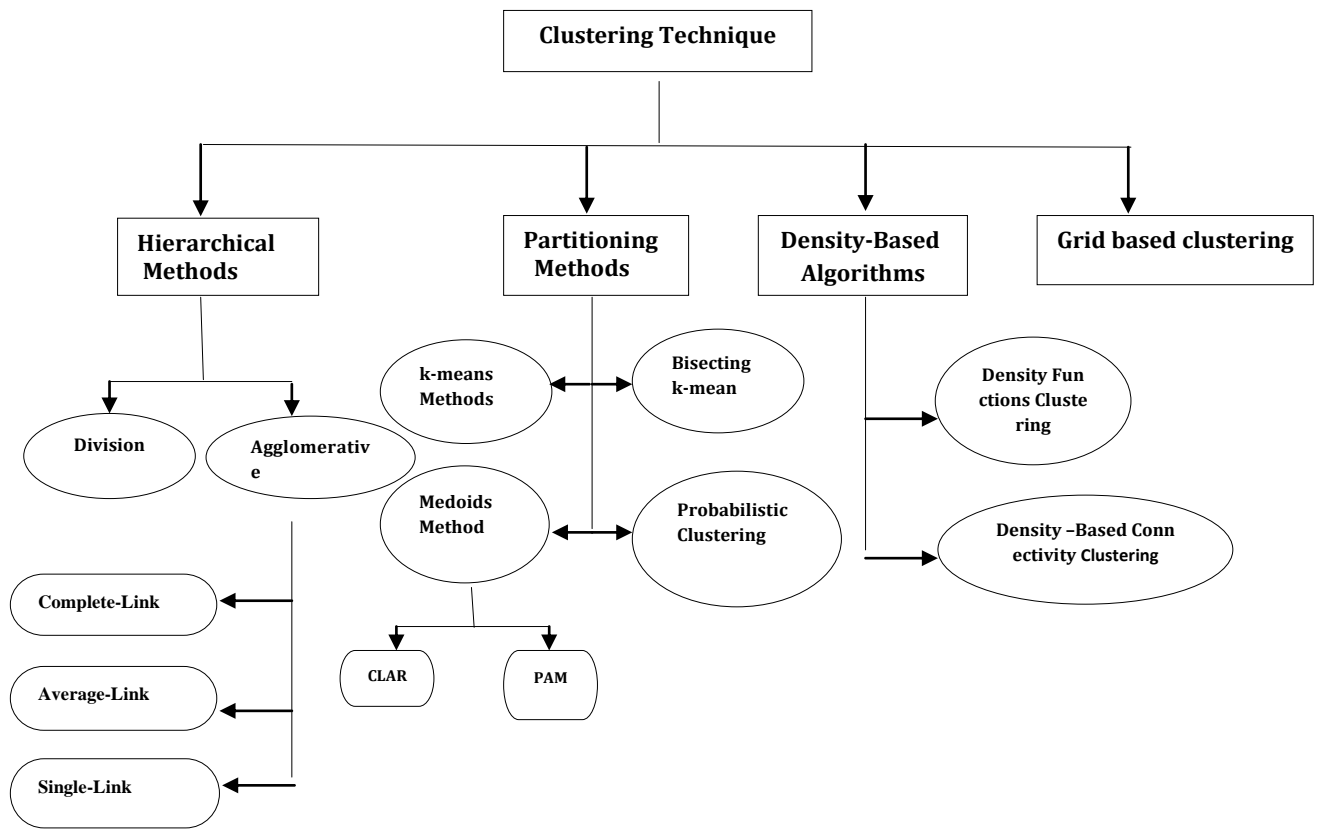

**Figure1** clustering methods division

## 2. Unsupervised classification algorithms

In this section, some clustering algorithms are studied. Summary of recent algorithms are presented in Table 1 with their advantages and disadvantages.

**Table 1** unsupervised classification methods

| Document clustering methods | Applied method | result | advantages | Disadvantages |
|---|---|---|---|---|
| [3] Congnan Luo et al (2009) | Selecting initial centroid based on candidate Combination of cosine function Increase of accuracy in document ranking and using new similarity | and link function as a similarity measure- new heuristic function based on neighbors of cluster centroid | selecting initial centroids k-means and bisecting k-means clustering algorithm | |
| | measure | for selecting a cluster to be divided in bisecting k-means algorithm | | |
| [4] M.Hanumanthappa et al (2012) | Algorithm change first the farthest points(M-FPF) | Producing cluster label | Improving the quality of clustering than FPF algorithms | Lack of improvement in computation time |
| [5] Yucong Liu et al (2011) | Representing document based on visual features and optimization of k-means algorithm based on mutation and crossover | Using mutation and crossover for improving k-means | Producing the value of k and cluster centroid dynamically | |
| [6],[7] VivekKumar Singh et al (2011),(2013) | Evaluating flat clustering algorithm such as k-means, heuristic k-means, fuzzy c-means with different representation Produce better results TF-IDF with all clustering method- the best clustering result with stemming or alone of (TF,TF-IDF, Boolean)and feature selection scheme( with or without stop words or with or without stemming) | | Better results of fuzzy clustering-FCM a stable and robust method and insensitive to initialization | |
| [8] Nadjet Kamel et al (2014) | Hybridization k-means and PSO algorithms using Improving initialization of k-means algorithms, sampling | improving PSO+k-Convergence to local optimum, least amount of fitness, having Increasing execution time, slow convergence means algorithm and useable for large dataset | compact and dense cluster | |

| Reference | Method | Description | Advantages | Disadvantages |
|---|---|---|---|---|
| [9] Jiayin Kang et al (2012) | Combination of PSO and FCM | Solving FCM local optimum, overcome to slow convergence of PSO algorithm | High value of F-measure and purity compared to FCM and PSO algorithms | |
| [10] ZonghuWang and et al (2011) | Initial centroids algorithm and partitioning algorithm | Producing initial centroids dynamically, constructing a method for optimizing initial centroid, improving the quality of cluster | Less chance for noise object to be selected as initial centroid, reducing the risk of getting stuck to local optimum | Determining the number of k manually, low accuracy in document clustering in interface of two cluster |
| [11] Md. Sohrab Mahmud and et al (2012) | Using average weighting value and ranking algorithm | Finding estimated centers by reducing the number of data iteration into the cluster, computational complexity less than k-means, better efficiency according to Producing initial centroids automatically, fast convergence to k-means algorithm, suitability for high dimensionality accuracy and time | | Need to provide desired number of clusters as input |
| [12] Murat Erisoglu and et al (2011) | Algorithm based on two main variable according to maximum variation coefficient and correlation value | Good performance for obtaining cluster centroids, better result compared to random initial centroids | Improving cluster structure and more adjustable than random method, easy implementation | Determining the number of cluster at first |
| [15] Kathiresan V and | Calculating cluster initial | Combination of | High accuracy | Predetermined k |
| et al (2012) | centroids k-means based on ranking method z-score and assigning data point to the clusters keeping information of distance to center of nearest cluster | systematic methods for finding initial centroids, produce initial centroid automatically, suitable for data with high dimension | efficiency, less time than k-means, suitable for document clustering | from beginning |
| [16]Shi Na and et al (2010) | Using two data structure Cluster[] and Dist[] for keeping information in each iteration and using them in each next Simple and efficient method for assigning data to cluster s, reducing computational Increasing the speed of cluster, reducing execution time, suitable for high dimension and Need to estimate the number cluster, selecting k objects from dataset as initial centers randomly, iteration to reduce calculation of distance between data point and cluster centroids | complexity O(nk), increasing efficiency | using it in text document | constructing local optimum |
| [17]RanjanaAgrawal and et al (2012 | | Optimal clustering of unknown text sets automatically, solving the mistake of zero cluster | Not need to k as input, working with empty clusters, working with lots of document in smaller time than k-means, F-measure better than k-means | |
| [18] RajeevKumar and et al (2012 | Include two phase. Initial clustering for finding k seeds as initial centroid and assignment of each document to k cluster. Second phase refining Constructing global optimum solution by minimizing criterion | | Minimizing the problem of getting stuck in local solution, high efficiency with increase of the number of cluster | |

| | function based on graph to achieve an accurate solution and using criterion function based on graph | | | |
|---|---|---|---|---|
| [19] Juntao Wang and et Fast global k-means and Constructing clustering Reducing sensitivity of kLimited to small dataset al (2011) | | LOF(Local Outlier Factor) | with high accuracy, increase of time with large dataset | -means to outlier |

The partitioning algorithms, such as k-means family has a good performance on document clustering. One of important characteristic of partitioning algorithms is that, it uses the global criterion function. The goal of this criterion function is to optimize different aspects of inter-cluster similarity and intra-cluster dissimilarity, and their combinations. Cosine function is known as a similarity criterion that is used widely in documents clustering algorithms. Cosine function can be used in the family of k-means algorithms to assign each document to a cluster with the most similar cluster centroid, in order to maximize the intra-cluster similarity. Since the cosine function measure the similarity of two documents, only the pairwise similarity is considered when it is determined whether a document is assigned to a cluster or not. When the clusters are not well separated, their partitioning only based on pairwise similarity is not good enough because some documents in different clusters may be similar to each other. To solve this problem Congnan Luo et al used the concepts of neighbors and link for document clustering. If two documents are similar enough, they are considered as neighbors of each other and link function between two documents show the number of their common neighbors. In [3] a method is proposed for selecting initial centroids of cluster which is based on three values; the pairwise similarity value calculated by the cosine function and the link function value and the number of neighbors of documents in the datasets. A new similarity criterion was proposed for determining the closest cluster centroid for each document during the cluster refinement phases. This similarity criterion is composed of the cosine and link functions. Also, a heuristic function was proposed based on the neighbors of centroid for the bisecting k-means to select cluster to split. Unlike the k-means algorithm, which split the whole dataset into k cluster at each iteration step, the bisecting k-means algorithm splits only one existing cluster into two sub-clusters, selecting of a cluster to split based on the neighbors of centroids instead of the size of cluster, improved accuracy of clustering, because the concept of neighbors provides more information about the intra-cluster similarity of each cluster and proposed method for selecting the cluster in bisecting k-means algorithm, a cluster whose centroid has the smallest number of local neighbors was proposed to split. The results show that the initial centroid selected in [3] was well distributed and each one is close to sufficient number of topically related documents, so that they improve the clustering accuracy.

Document clustering is an effective tool for managing information overload. Grouping similar documents together enable us to have an Observer human to quickly review large document classification which makes it possible to easily understand the distinct title and subtitle. [4] Is dealt with text document clustering using modified FPF[1] algorithm and comparison of it with FPF clustering algorithm. Search engines are very invaluable tool for retrieving information form the web. The search engines prepare a list of ranked results according to query posed by the user. But the main problem is that user can't really find what is related to his information. Due to the fact that query may be short and its statement might be ambiguous. In contrast, there are some methods for retrieving web information based on categorization and they allow the user to observe the results related categories that has better matching with user query, but only covers a small fraction of existing pages. Now the main method is search results clustering. Clustering avoids from demonstrating some equivalent documents in the first page of results and this is one of advantages of clustered search engines. FPF includes two phases of initialization and iteration In initialization phase an arbitrary point is selected as head1 (the head of cluster 1) and all points are assigned to cluster 1. In second phase at each of iteration of algorithm the furthest point form the current set of heads selected as $head_i$ and every point closer to $head_i$ moves to cluster *i* form its current head. [4] With M-FPF algorithm and using random sampling improved output clustering quality, but there is not any changes in calculation in M-FPF algorithm. FPF works as a search engine that groups into disjoint labeled clusters the web snippets returned by auxiliary search engines. FPF works in three phases: 1) extracting keywords of cluster 2) selecting keywords 3) generating cluster label. Generating label by M-FPF provides user a compact guide to recognize relevance of each cluster to the required information.

---

[1] Furthest Point First

[5] Proposes a method to cluster the scientific documents based on visual features (VF). Five of visual features of document include title, keywords, subtitle, abstract, body. In this method, mutation and cross over in genetic algorithm is used to improve $k$-mean algorithm by adding and integrating cluster centroid during clustering process to adjust the value of $k$ and cluster center dynamically. The proposed algorithm [5] works in several stages. At first, initialization the cluster centroids Using similarity threshold $\lambda$, then calculating the similarity Between each data point and each cluster center, and comparing the biggest similarity with $\lambda$. If the similarity is bigger, the data put into cluster with biggest similarity and data should be added into cluster center using mutation, which can cause change in the number of cluster center. In next step, the center of each cluster is obtained by averaging all of its data are calculated again. After that, the similarity of each pair of cluster centers which is obtained in previous step is calculated. In this step if the similarity between two clusters is bigger than $\lambda$ then they have to be merged. These steps are repeated until that cluster center reaches a stable value. According to the obtained results, accuracy and stability of clustering subtitle only is less than the body, but due to the very little size of subtitle, its efficiency is much better than the body. If the efficiency is a basic factor, cluster by keywords and subtitle is the optimal choice.

Clustering is the same as classification, but they are different in that classes are not pre-defined and process of categorization is unsupervised. Clustered objects are grouped based on the principle of maximum of similarity among the members of each class and minimum of similarity among different classes. It means that cluster is adjusted in a way that members of each cluster have the most similarity together. From one aspect, clustering algorithm is classified to two classes: hard and soft. In hard clustering, each data point belongs to one cluster, while in soft clustering each of data point with a particular membership value is assigned to each cluster and this value is in the interval [0, 1]. K-means is a hard clustering and FCM is a kind is a kind of soft clustering. VivekKumar.s et al [6] and [7] conducted some experiments for evaluating some flat, soft and hard clustering algorithms such as k-means, heuristic k-means and fuzzy c-means with different representation(TF,TF-IDF, Boolean) and different feature selection scheme( with or without word stop, word removal and with or without stemming) on some standard datasets. The results show that TF-IDF representation and use of stemming scheme obtains better results with all clustering methods of different datasets and also TF representation is marginally better than Boolean but worse than TF-IDF. Although Boolean presentation produces relatively good results for classification task but it is not a suitable choice for clustering. Stemming produces better results for clustering. Removing stop words, though improves the quality of clustering in some cases but totally its effect is not significant. Surprisingly, stemming and removing stop words together does not produce better results compared to stemming alone. Fuzzy clustering produces better results than k-means and heuristic k-means an all datasets. FCM is a stable and robust method that is quit insensitive to initialization and it converges less to the local minimum.

As pointed earlier, this paper focuses on partitioning clustering. In this section, some works performed on solving the problem of partitioning algorithms such as k-means are presented. But these methods almost can solve some of the problems and finally all of them try to improve clustering accuracy.

**2.1. Dependency of final results to initial centroids initialization**

Hybrid clustering is one of conventional methods in data mining that is obtained from hybridization of initial clustering methods and results in improvement of clustering accuracy than original methods. [8] Uses k-means and PSO algorithms to solve the problem of initialization of k-means algorithm of hybrid algorithms, using sampling algorithm to improve the results of clustering. Particle Swarm Optimization (PSO) is a creative optimization method based on probability rules. PSO technique is mainly used for optimization problem. Using hybrid methods solves the problem of clustering includes need to determine the number of cluster but their problem is slow convergence. The problem of document clustering is their multi-dimensionality that can be overcome by the hybrid method. PSO algorithm, alone has a problem of convergence slowly. In hybrid model of k-means and PSO, PSO algorithms is used to produce initial centroid for k-means algorithms, then the results of centroids are used by k-means algorithm. In PSO+k-means algorithm, global search of PSO and rapidity of convergence of k-means are combined. Refining the algorithm of the initialization by sampling is also a solution for initialization problem of k-means algorithm which is sampling a collection of data in a set of sub-groups and using k-means algorithm on each group. Sampling method is efficient on the dataset with large size. The proposed algorithm [8] is such that S sub-sample are selected from initial set, (number of particles) then initial centroids is constructed randomly by implementing k-means algorithm so a set of centroids is obtained for each sub-sample and PSO algorithm which consists of S particles implemented on results obtained in previous step, finally k-means algorithm is implemented on it. The main purpose is constructing a good initial structure for PSO algorithm and reducing the number of particles. PSO requires a big number of iteration and particles because its initialization is random and in the case of a large collection of data, complexity is a main problem and use of PSO+k-means algorithm for clustering needs large memory and high implementation time.

FCM is a strong unsupervised clustering that widely used for solving the problems of clustering. Although, implementing FCM is easy and it work fast in most of conditions, but it is sensitive to initialization and convergence to local optimum. In [9] a hybrid method is proposed for text document clustering based on FCM and Particle Swarm Optimization (PSO), so that FCM does not lead to optimum. This algorithm is a combination of the merits of both algorithms and overcomes to slow speed of convergence in PSO algorithm. The proposed algorithm uses the ability of PSO algorithm in global searching to overcome the problems of FCM. The performance of this hybrid method PSO+FCM is better than traditional algorithms of PSO and FCM.

Most partitioning algorithms are sensitive the centroids and the results of clustering mainly depend on the initial centroids. Zonghu Wang et al [10] applies unsupervised feature selection method for reducing dimension of document feature space and proposes a new algorithm based on partitioning which selects initial cluster centroids in the process of clustering with respect to the size and density of cluster in the datasets. the algorithm has three steps: 1) generating k density groups as initial centroids set 2) to assign the remaining documents to their nearest groups by measuring similarity between documents and clusters 3) to refine the results of clustering by reassigning documents to their nearest groups. The results of several text datasets indicate that the proposed algorithm improves the quality of cluster effectively. Usually in k-means clustering method initial centroids are determined randomly, therefore determining centroids causes to obtain nearest local minimum and not global optimum. The method explained in [10] proposes in order to optimize the initial centroids for partitioning the datasets which produced by a combination of documents pair similarity measure and the similarity of candidate centroids of generated clusters. By this approach, the noisy objects have little chance to be selected as initial centroids and dense and large clusters will be selected earlier, so the chance of getting stuck in poor local optima is reduced. Using this method can partition the documents which are at the boundary of two clusters reasonably.at first some document may be hard to be partitioned; so that the size of initial centroids increases these indistinguishable documents may find their nearest clusters.

As mentioned earlier, the classic k-means algorithm strongly depends on the selection of initial centroids, then accuracy and quality of cluster deeply depends on selected initial centroids. Several methods are proposed for improving the efficiency and accuracy of k-means clustering algorithm which show that initial centroids better can improve accuracy and efficiency of clustering groups and also improve complexity time. In order to overcome the problem of k-means algorithm for selecting initial seed point (clusters), [11] proposes a heuristic method to find better initial centroids and more accurate cluster with less computational time. In improved algorithm, an uniform method is used to find rank score by averaging the feature of each data point which will produce initial centroids that follow the data distribution of the given set. This method uses a sorting algorithm to the score of each data point and data is divided into k subset in which k is the number of desired clusters. Then, the mean value from each subset is calculated and the nearest value of mean from each subset is considered as initial centroid. The complexity of this algorithm is $O(n(kl+logn))$. Therefore, due to independency of this algorithm to dimension, it is suitable for text documents with high dimensions. The results show that the algorithm [11] produces clusters with better accuracy and improve the effectiveness of k-means algorithm, production of initial centroids are computed automatically and optimal centroids are found by the program, while in original performed randomly. But this algorithm has one limitation that the number of desired cluster should be determined as input.

[12] Proposed an algorithm for calculating initial cluster center for k-means algorithm in which the structure of clusters has been improved and is compatible with random initialization. In this algorithm, two main variables are selected according to maximum variation coefficient for main axis and minimum absolute value of correlation between selected variable for main axis and other variable for second axis. After determining both axes, according to selected axis, the mean of data point as center of dataset is calculated and Euclidian distances between each data point and the center ($d_{im}$) is computed, then the point that has highest distance is selected as the first candidate from the center of initial cluster ($c_1$), then the distance between data points and $c_1$ is calculated ($d_{c1}$). In order to select a candidate for the second initial cluster center the same mechanism is applied and $d_{jc1}$ instead of $d_{im}$ and also the data point which has highest distance of $d_{c1}$ is selected as the second candidate of initial cluster center ($c_2$). This process is repeated until the numbers of initial cluster center equals to predefined number of clusters, then cluster membership of each point are

determined according to candidate initial cluster centers and selected two axes. This method compared to CCIA [13] algorithm and proposed method of Deeler [14] is better according to error percentages and almost all clusters include data.

In [15] a method is proposed based on z-score for calculating initial cluster center for k-means algorithm. Z-score is a statistical method of ranking numerical and nominal attributes based on distance measure. Data is ordered based on score value then the ranked data is divided into k subsets. The mean value of each subset is calculated and the neatest data value is selected as initial centroid. The method [15] include two phases: the first phase is to determine initial centroids using z-score method and the second phase is to assign data points to the nearest cluster which is a repetitive process using heuristic method for improving efficiency. In second phase, produced initial centroid in previous step is considered as inputs. Then distance between each data-point and initial centroids of all clusters is determined and this data-point is assigned to cluster centroid. For each data-point the cluster to which it is assigned (Cluster ID) and its distance from the centroid of the nearest cluster (Nearest Dist) are determined. For each cluster the centroid are calculated by taking the mean of new centroid. This step is similar to the original k-means algorithm except that the initial centroids are computed automatically. In next step, by an iterative process distance between each data-point from the nearest cluster centroid is maintained and at the beginning of iteration the distance from the point to new centroid of the nearest cluster is calculated. If the distance between data-point and the new centroid of nearest cluster is less than or equal to the previous nearest distance, then that data-point stays in its cluster and there is some saving of time in calculating the distance between data-point to other k-1 cluster centroid. Otherwise, it is possible that the data-point is assigned to the other nearest cluster, so the distance between the data-point to all cluster centroid is computed and recorded in the nearest cluster and new distance value is stored. This loop is repeated until there is no more point in cluster boundaries and the convergence criterion is determined. The heuristic method in [15] reduces calculation significantly and thus improves the efficiency and this reduction in the number of calculations makes algorithm to implement faster in liner time, so it is suitable for document clustering but the number of cluster still should be specified.

Since the k-means algorithm takes a lot of time for executing, especially for datasets with large capacity, then [16] present a simple and efficient method for this shortcomings of k-means algorithm that improves clustering speed effectively and reduces the complexity of computations $O(nk)$. In this method, the same as second step, assignment of data-point to the cluster is similar to the process mentioned in [15] and two structures of the data is used for keeping cluster label and the distance of all subject to the nearest cluster and it has a similar iterative process. With respect to time complexity, if the data-point remains in the original cluster this needs O(1) and otherwise needs O(k). If half of the data-point move from their cluster the complexity is O and totally is O(nk). The information in data structure allow this function to reduce the number of distance calculation required to assign each data object to the nearest cluster and this method improves k-means algorithm and makes its execution faster. But still there is problem with determining the number of cluster k and several k should be tried to reach a desired solution. Since this algorithm selects k objects from dataset as initial centroid, then this method is sensitive to initial centroid. Therefore, it does not produce consistent clustering results duo to random selection of initial points. So, it is possible to obtain local optimum for determining the centroids.

## 2.2. Determining the number of cluster *k*

Partitioning algorithms have some limitations such as the number of clusters has to be given as input, they converge to different local minima based on the initializations and they cannot work with empty clusters. Working with partitioning algorithm using cosine distance for text data is another limitation and if the document has value zero in vector space model, it does not get clustered and partitioning algorithm stops clustering. To solve these problems, [17] proposes a new algorithm to produce number of clusters automatically and it clusters unknown text corpora in optimum clusters, then it does not take input as number of clusters. K-means algorithm cannot work well with large number of documents due to zero error (zero value of VSM and cosine similarity $\infty$). The proposed algorithm solves this problem. Limitation of the algorithm is when documents are clustered in one category. The proposed algorithm is compared with k-means algorithm on different dataset and the results show that the value of F-measure of proposed algorithm is better than k-means algorithm.

## 2.3. Convergence to local optimum

Most of the used partitioning clustering algorithms are inflicted with the drawback of trapping into local optimum solutions. Rajeev.k et al [18] proposed a partitioning algorithm for document clustering using a graph based criterion function that usually leads to the global optimum solution and its performance enhances with increase in the number of clusters. Unlike other clustering algorithms such as k-means, the proposed algorithm uses a greedy method. In the first phase, the purpose is obtaining k seeds each representing a single cluster and second phase involves the repeated change in the structure of clusters. This algorithm wisely minimizes the problem of getting stuck in any local solution and it successfully obtains the desired global solution. But due to this fact that it checks all possible solutions before reaching the global optimal solution, its execution time is increased. The presented method is compared with several known algorithm such as k-means and k-means++ and the results showed that in the proposed algorithm, entropy decreases with increase of the number of clusters and the value of F-measure increases.

**2.4 Sensitivity to noisy and outlier points**

K-means algorithm is an algorithm with high efficiency, scalability and fast convergence at the time of dealing with large datasets. But it has many shortcomings includes that the number of clusters k needs to be initialized and initial cluster centroids are selected arbitrarily and also the algorithm is affected by noisy points. Considering these limitation of traditional k-means clustering algorithm, Juntao wang et al [19] propose an improved k-means algorithm using noisy data filtering. K-means algorithm is very sensitive to initial cluster centroids, so that the clustering results is very different from different initial cluster centroids. If the data-points exist in isolation, that is a small amount of data-points are far from data-intensive areas, the calculation of mean point would be affected, and the new cluster centroids may deviate from the true data-intensive area which eventually lead to a clustering output result of large deviation. [19] Removes the isolated points from datasets before clustering using outlier detection method based on LOF[2], so that the outliers cannot participate in the calculation of the initial cluster centroids and excluded the interference of outliers in the search for next cluster center point. In second stage, Fast Global k-means is applied on searching for the next cluster centroids points. According to the improved algorithm, as isolated points are excluded from the data set with applying detection of density based outlier so that the sample which is farthest from the cluster center can be selects as the next best initial cluster center, the sensitivity to outliers in k-means algorithm is decreased effectively and accuracy of clustering is improved. This algorithm is suitable for small datasets and cost more time when dealing with large datasets.

**3. Supervised classification algorithms**

This section is dealt with classification algorithms. Table2 present a summary of studied

---

[2] Local Outlier Factor

algorithms.

Table2-Summary of supervised classification methods

| Supervised classification methods | Applied method | result | advantages | disadvantages |
|---|---|---|---|---|
| [20] .Borodin and et al(2013) | Learning from misclassified documents | Using cluster head result in uniform data and Not have over-fitting and over-smoothing adjust with every change in document distribution | | High computational complexity |
| [21] .Ko and et al(2009 | Using bootstrapping and feature A classification with reasonable accuracy can Not need to labeled documents projection | achieved only by subtitle words and unlabeled data | | Existing noisy Wrong labeled document, dependency to quality of keyword |
| [22] P-Y.Hao and et al(2007 | Using SVM hierarchically | Using hierarchical structure causes increase of accuracy | Higher accuracy, more flexibility | High computational complexity |
| [23] C.Jiang and et al(2010) | Documents classification using weighted graphs | Using weighted sub-graphs result in pruning unuseful nodes | Increasing accuracy, reducing search space | High computational complexity |

| | | and reducing search space | | |
|---|---|---|---|---|
| [24] S.Chaghari and et al(2011) | Classification based on content and structure | Using structure and content give better results than classification based on Increasing accuracy content | | High computational complexity |
| [25] N.Tsimboukakis and et al(2011) | Two stage classification based on content | | Increasing accuracy | Ignoring the structure |
| [26] C.H.Wan and et al(2012) | Combination of K-NN method and vector space model | Using SVM in training stage can increase accuracy and reduce dependency to the neighbors number | Non-dependency to the number of neighbors | High computational complexity |
| [27] S.Jiang and et al(2012) | Improving K-NN method | With Combining clustering in training stage in K-NN method can achieve to better efficiency | Reducing the computation of text similarity, more efficiency | Sensitivity to noisy and outlier |
| [28] X-g.Zhao Using ELM with Using v-ELM | voting to classes | give better results | accuracy, faster | hidden node layer |

| Improve more and et al(2011) | Having | than ELM | learning, less training time, less complexity of neural network | |
|---|---|---|---|---|
| [29] A.Gordo and et al(2012) | Embedding projection vector into common sub-space of cheap and expensive features | Using CCA and cheap feature can increase classification accuracy | Not need to expensive features | Limited to vector data |
| [30] E.K.Aydongan and et al(2012) | Combining genetic algorithm in fuzzy rules | Using steady-state genetic algorithm has higher accuracy than genetic algorithm | Increasing accuracy, less time | Non-convergence to optimum result |

Many of supervised and unsupervised classification methods use the center as class representative for document classification. Due to using centroids for classification, over-smoothing and making the class representative stick to the document center of mass. Since these classification algorithms are centroid-base, then they need to be applied on input data sets periodically. In [20] an algorithm is proposed for documents classification that is called 3LM[3]. 3LM builds on the ideas of supervised centroids-based documents classification and K-means unsupervised clustering algorithm. In this algorithm, cluster heads are used instead of centroids, and this allows 3LM to fit better for distributing data. In this method, after classifying documents into predefined classes, cluster heads are determined. The cluster head is a vector representative of a class with is incrementally update to be closer to the misclassified document of its. This algorithm tries to pull cluster head toward misclassified document in order to better distributing. In such a way that in it each stage, a bisector is drawn between the pair of document and the cluster head is directed to the bisector. This algorithm can adjust with every change of document distribution due to change in position of cluster head. 3LM uses misclassified document also it uses negative feedback to improve the classification accuracy and to create a better fit for distribution of the data. Also this algorithm never stops learning and it uses a fixed learning rate. The proposed method prevents over-fitting and over-smoothing due to use of cluster head. This is occurred when a model is too complex and has too much parameter than the number of data which make model to report noise or random error instead of the original relationship.

---

[3] Life Long Learning form Mistakes

The main problem with supervised learning algorithms is that they require a large number of labeled training documents for accurate learning. In [21] a new text classification method based on unsupervised or semi-supervised learning is proposed. This method uses only unlabeled documents and the title word of each category as initial data for learning of text classification. Also, this method applies an automatic labeling task using a bootstrapping technique. The main and final goal of bootstrapping process is to build up labeled training documents from only title words automatically. This process includes three phase: first, A set of content vectors is constructed which represent content of each context. In the phase context- clusters is constructed for training. In this way, the keyword which related to title world semantically are collected and a context with a keyword or a title word of any category is selected as a centroid-context and they are sorted in a descending order according to their calculated scores. Finally, the context-cluster of each category is constructed by measuring similarity between word and context using two matrixes WSM[4] and CSM[5]. Each category has one WSM and one CSM. The row and column of WSM are labeled by all the existing content words in centroid-context of each category and input remaining contexts. The rows of CSM include centroid-contexts and columns to the remaining contexts. Then affinity between words and context and also similarity between word and average of affinity to content between every pair words are calculated. Finally, Naïve Bayes is used for learning from context-clusters. Since data is obtained by bootstrapping method, then they include more incorrectly label documents than manually labeled data. Therefore, TCFP classifier with robustness from noisy data is used.

One of main problem with classification method is that they ignore the hierarchal structure and their classification structure is flat so that the pre-defined categories are treated separately and the structure of relationships is not defined. While classification based on structure and content increases the accuracy more than classification based on content. In [22] a classification method is offered that has more flexibility in designing classifier system. In hierarchical method, each node is divided to some smaller sub-nodes for which are suitable classifier or a technique of feature extraction is used that determine input data range. In the proposed method, after selecting features, the hierarchical structure is constructed using unsupervised support vector clustering method. Then, in the unsupervised learning stage, SVM[6] classifier is used as tree-node. This method has higher accuracy compared to non-hierarchical SVM documents classification system such as K-NN[7] and decision tree.

---

[4] Word Similarity Matrix
[5] Context Similarity Matrix

Most of document classification formalization is the vector space model founded on the bag of words/phrases representation. The main advantage of the vector space model is that it applies on classification algorithm simply. In this method structure and semantic of information is ignored, while structural information has an important rate in accuracy of classification, so graph is used for applying structural role in document classification. Of course using graph has higher level of computational complexity compared to the vector. In most of graph representation, it is assumed that sub-graph have equal importance. In [23] proposes a method for documents classification based on weighted sub-graphs that consider different significance for sub-graphs. In the algorithm of frequent sub-graph mining process, it could be possible to reduce the search space. In the proposed algorithm, after pruning unuseful sub-trees by weighted-gSpan, the weights of remaining sub-graph which are more significant are calculated using one of three methods includes structure based, content based and structure and content based methods then it is used as input for classification.

Also, another method is proposed in [24] that classify XML documents with representing structure and content of the document using the vector model. In this method leaves are the text of documents and node are tags or documents structure. In order to reduce the computational complexity, the nodes with the same label and structure merge at the same depth. In this way, an aggregated tree is constructed. After that, the key terms of each node include nouns and verbs are extracted. In proposed algorithm, each feature is represented as label-term. Then weight of term is calculated using TF-IDF formula and eventually, SVM is used for learning and classification. It is observed that using structure and content in classification improved metrics such as recall precision and F-measure better than classification based on content and therefore it improves accuracy.

---

[6] Support Vector Model
[7] K-Nearest Neighbor

Also in [25] a classification system is proposed based on content. In this two-stage system, where only the second one is supervised, at first a word map is created to represent features of the document in the pattern space. Then in second stage a supervised classifier is applied. Two unsupervised alternative are proposed for creating word map in the first stage: the first one is based on SOM[8] model and employs a rule based on Pareto analysis to reduce the dimensionality of content-representation vectors which in order to compress further, k-means algorithm is used. The second one is based on HMM[9] which is used for content coding. HMM implements an unsupervised fuzzy clustering of words. Obtained results show that proposed method has higher accuracy than SVM. Only when the number of SVM features is increased, SVM accuracy will be higher than the proposed algorithm that is impractical in real world.

K-NN and SVM classification methods are widely used in several kind of documents classification. K-NN is applied due to its low implementation cost and high degree of effectiveness, but the accuracy of classification decreases when the value of k increases until a certain threshold. The reason for this fact is that too many neighbors leads to occurrence of noisy. SVM classifier is used in many classification methods due to its high accuracy. In proposed method [26], SVM is used for reducing training samples to their support vectors from the nearest data points to separating plain. In this approach called SVM-NN[10], at first SVM training is used to reduce the training data points to the support vectors (SVs). Actually, SVS are neighbors of test document which are used instead of k. Then Euclidean distance function is used to calculate distance between input data point and SVs of different categories. This method is not dependent deeply on the number of neighbors compared to K-NN and even it can ignore K-NN in training stage, especially when the training samples are limited and insufficient for preparing training set.

When traditional text classification algorithms face to imbalance documents their performance decreases greatly. Among existing algorithms, the robustness of KNN and SVM is better than Naïve Bayes because of its adaptive weight based on the number and distribution of training text samples.as mentioned K-NN algorithm has three main defects: first, the complexity of its sample similarity computing is huge. Second, its performance is easily affected by noisy sample. Third, it is a lazy learning method and does not build the classification model. So, it is not suitable for many applications in which data is updated dynamically and required deeply in real-time performance. In [27] an improved K-NN algorithm is presented. This algorithm is called INNTC[11] which builds the classification model using constrained one pass clustering algorithm. One pass clustering algorithm evaluate the text only in one pass and uses the least distance principle to divide training text samples into hyper-plain with the same radius. It change the learning way of K-NN algorithm and the number of clusters obtained by constrained clustering is much less than the number of training samples. VSM[12] algorithm is used to represent the documents. In this model each document is considered as a vector in the term-space. After obtaining the classification model, K-NN is used to classify the test text sets. As a result, when K-NN method is used for classifying, similarity computing complexity is decreased significantly.

---

[8] Self-Organizing Map
[9] Hidden Markov Model
[10] SVM-Nearest Neighbor

Another algorithm used for documents classification is ELM[13]. In ELM hidden node parameters are selected randomly and it usually requires much less training time than the conventional learning machines. In [28] an improved ELM algorithm called voting-ELM is proposed for classification of XML documents. In the proposed algorithm, m classes are decomposed into pair. Each pair includes one ELM classifier. In v-ELM each classifier has a lower neural network complexity and a shorter training time. Also, v-ELM has more hidden layer nodes than ELM. After applying v-ELM classification on documents, two post-processing stages are also proposed. If two documents have equal maximum votes, then voting process is repeated again. Also, if maximum difference value of votes of two classes which has highest votes to an especial class and minimum difference value between second and third maximum is equal to defined threshold, then voting process is repeated to improve accuracy of classification. V-ELM method has higher learning speed and due to decompose classes and repeating voting process and also accuracy more than ELM.

In [29] features used for representing documents divided to two categories: features that are available in training time are called expensive features and those are available in both training and test times are called cheap features. In this method classification of documents is performed by applying analysis CCA[14] using cheap features based on expensive features. Then, a common subspace is found between cheap and expensive features and a set of vectors embeds features into that subspace. After that, cheap features can be projected into that common subspace and be used to classifier. At test time cheap features project into the subspace and classifier is used without any need to access to expensive features. In this way, CCA can improve the accuracy of documents that some of their features are available only in test time, but this method is limited to vector data.

---
[11] Improved K-NN algorithm for Text Categorization
[12] Vector Space Model
[13] Extreme Learning Machine
[14] Canonical Correlation Analysis

In [30] a hybrid genetic algorithm (HGA) in fuzzy rules is presented for documents classification. A fuzzy rule-based classification system (FRBCS) includes two main components: knowledge based (KB) and fuzzy reasoning method (FRM). KB has two components: data based (DB) and rule based (RB). DB contains the labels used for variables and RB consists of a set of rule used for classification and a mechanism for documents classification using DB information based on the rules. In the proposed algorithms after determining the rules, HGA is used to extract the classification rules. In this algorithm each individual symbolized a rule for each class and genetic algorithm is used for producing several rules for each class and eventually IPF[15] is used for selecting rules among set of rules. In proposed algorithm, steady-state genetic algorithm is used. This algorithm is different from traditional genetic algorithm in which a single new member inserted into the population and generally the worst individual is removed from the population. Additionally, consumed time in this algorithm is much smaller and its accuracy is higher than traditional genetic algorithm. Also, this algorithm allows some input variables in each rule to be absent. In each generation of HGA, a heuristic procedure is implemented as a local search to improve the offspring obtained by genetic operators In this procedure, the value of a gen of chromosome is changed randomly.. If the new result is better than previous one, it is replaced. Finally, after a predetermined number of iteration, the best result is selected as offspring. This algorithm has higher accuracy and smaller consumed time than traditional genetic algorithm.

## 4. Conclusion

Many researches are conducted in the field of text documents classification and this subjects is still attracts the researchers' attentions. Text classification algorithms try to classify the texts based on their similarities and put the related texts into the some group. Documents classification is used in applications such as retrieving information, summarizing the texts, extracting the keywords and organizing the documents etc. Several methods are proposed to classify input dataset which aim to achieve to desired number of classes with maximum similarity between the elements of inside of the category and minimum similarity with other categories. Totally, all classification methods have advantages and disadvantages. Methods using graph instead of vector model despite of having higher accuracy suffer from computational complexity. Also, supervised classification algorithms need a lot of labeled training documents for accurate learning which require more computational time than unsupervised classification algorithms. Also, the methods deal with structure and content simultaneously have higher accuracy and more calculations than methods using structure or content alone. Therefore, depending on importance of documents and cost of time and consumed memory and existing conditions, a suitable classification method has to be selected.

---

[15] Integer-Programming Formulation

**Refereces**

[1] P. Rai and S. Singh, "A Survey of Clustering Techniques," vol. 7, October 2010.

[2] M. Bakhshi, M.-R. Feizi-Derakhshi, and E. Zafarani, "Review and Comparison between Clustering Algorithms with Duplicate Entities Detection Purpose," *Computer Science & Emerging Technologies,* vol. 3, June 2012.

[3] C. Luo, Y. Li, and S. M.Chung, "Text document clustering based on neighbors," *Elsevier,* vol. 68, pp. 1271-1288, 2009.

[4] M.Hanumanthappa, B.R.Prakash, and M.Mamatha, "Improving the Efficiency of Document Clustering and Labeling Using Mudified FPF Algorithm," in *AISC*, India, 2012, pp. 957-966.

[5] Y. Liu, B. Zhang, K. Xing, and B. Zhou, "Document Clustering Method Based on Visual Features," 2011.

[6] V. K. Singh, T. J. Siddiqui, and M. K. Singh, "Evaluating Hard and Soft Flat-Clustering Alghorithms for Text Documents," in *Advacnces Intelligent Human Computer Interaction*, Verlag Berlin Heidelberg, 2013.

[7] V. K. Singh, N. Tiwari, and S. Garg, "Document Clustering using K-means, Heuristic K-means and Fuzzy C-means " presented at the Computational Intelligence and Communication Systems, 2011.

[8] N. Kamel, I. Ouchen, and K. Baali, "A Sampling-PSO-K-means Algorithm for Document Clustering," presented at the Advances in Intelligent Systems and Computing, 2014.

[9] J. Kang and W. Zhang, "Combination of Fuzzy C-Means and Particle Swarm Optimization for Text Document Clustering," presented at the Advances in Electronical Engineering and Automation, Verlag Berlin Heidelberg, 2012.

[10] Z. Wang, Z. Liu, D. Chen, and K. Tang, "A New Partitioning Based Algorithm For Document Clustering," presented at the Fuzzy Systems and Knowledge Discovery (FSKD), 2011.

[11] S. Mahmud, M. Rahman, and N. Akhtar, "Improvement of K-means Clustering algorithm with better initial centroids based on weighted average," presented at the Electronical and Computer Engineering, Dhaka,Bangladesh, 2012.

[12] M. Erisoglu, N. Calis, and S. Sakallioglu, "A new algorithm for initial cluster centers in k-means algorithm," *Elsevier,* vol. 32, pp. 1701-1705, 2011.

[13] S. S.Khan and A. Ahmad, "Cluster center initialization algorithm for K-means

clustering," *Elsevier,* vol. 25, pp. 1293-1302, 2004.

[14]  S.Deelers and S.Auwatanamongkol, "Enhancing K-Means Algorithm with Initial Cluster Centers Derived from Data Partitioning Along the Data Axis with the Highest Variance," *Electronical and Computer Engineering,* 2007.

[15]  K. V and P. Sumathi, "An Efficient Clustering Algorithm based on Z-Score Ranking method," presented at the Computer Communication and Informatics, INDIA, 2012.

[16]  S. Na and L. Xumin, "An Improved k-means Clustering Algorithm," presented at the Intelligent Information Technology and Security Informatics, 2010.

[17]  R. Agrawal and M. Phatak, "A Novel Algorithm for Automatic Document Clustering," 2012.

[18]  R. Kumar, A. Ranjan, and J. Dhar, "A Fast and Effective Partitioning Algorithm for Document Clustering," Verlag Berlin Heidelberg, 2012.

[19]  J. Wang and X. Su, "An improved K-Means clustering algorithm," 2011.

[20]  Y. Borodin, V. Polishchuk, J. Mahmud, I. V. Ramakrishnan, and A. Stent, "Live and learn from mistakes: A lightweight system for document classification," *Elsevier,* vol. 49, pp. 83-98, 2013.

[21]  Y. Ko and J. Seo, "Text classification from unlabeled documents with bootstrapping and feature projection techniques," *Elsevier,* vol. 45, pp. 70-83, 2009.

[22]  P.-Y. Hao, J.-H. Chiang, and Y.-K. Tu, "Hierarchically SVM classification based on support vector clustering method and its application to document categorization," *Elsevier,* vol. 33, pp. 627-635, 2007.

[23]  C. Jiang, F. Coenen, R. Sanderson, and M. Zito, "Text classification using graph mining-based feature extraction," *Elsevier,* vol. 23, pp. 302-308, 2010.

[24]  S. Chagheri, C. Roussey, S. Calabretto, and C. Dumoulin, "Technical Documents Classification," in *Computer Supported Cooperative Work in Design*, IEEE, 2011.

[25]  N. Tsimboukakis and G. Tambouratzis, "Word-Map System for Content-Based Document Classification," *IEEE,* vol. 41, pp. 662-672, SEPTEMBER 2011.

[26]  C. H. Wana, L. H. Lee, R. Rajkumar, and D. Isa, "A hybrid text classification approach with low dependency on parameter by integrating K-nearest neighbor and support vector machine," *Elsevier,* vol. 39, pp. 11880-11888, 2012.

[27]  S. Jiang, G. Pang, M. Wu, and L. Kuang, "An improved K-nearest-neighbor algorithm for text categorization," *Elsevier,* vol. 39, pp. 1503-1509, 2012.

[28] X.-g. Zhao, G. Wang, X. Bi, P. Gong, and Y. Zhao, "XML document classification based on ELM," *Elsevier,* vol. 74, pp. 2444-2451, 2011.

[29] A. Gordo, F. Perronnin, and E. Valveny, "Document Classification Using Multiple Views," presented at the IAPR International Workshop on Document Analysis Systems, IEEE, 2012.

[30] E. K. Aydogana, I. Karaoglanb, and P. M. Pardalosc, "hGA: Hybrid genetic algorithm in fuzzy rule-based classification systems for high-dimensional problems," *Elsevier,* vol. 12, pp. 800-806, 2012.